\documentclass[aps,english,showpacs,twocolumn,pr]{revtex4-1}
\usepackage{amsfonts}
\usepackage{amssymb}
\usepackage{amsmath}
\usepackage{graphicx}
\usepackage{epsfig}

\begin{document}

\title{Inversion symmetric non-Hermitian Chern insulator}
\author{H. C. Wu}
\author{L. Jin}
\email{jinliang@nankai.edu.cn}
\author{Z. Song}
\affiliation{School of Physics, Nankai University, Tianjin 300071, China}

\begin{abstract}
We propose a two-dimensional non-Hermitian Chern insulator with inversion
symmetry, which is anisotropic and has staggered gain and loss in both $x$
and $y$ directions. In this system, conventional bulk-boundary
correspondence holds. The Chern number is a topological invariant that
accurately predicts the topological phase transition and the existence of
helical edge states in the topologically nontrivial gapped phase. In the
gapless phase, the band touching points are isolated and protected by the
symmetry. The degenerate points alter the system topology, and
the exceptional points can destroy the existence of helical
edge states. Topologically protected helical edge states exist in the
gapless phase for the system under open boundary condition in one direction,
which are predicted by the winding number associated with the vector field
of average values of Pauli matrices. The winding number also identifies the
detaching points between the edge states and the bulk states in the energy
bands. The non-Hermiticity also supports a topological phase with zero Chern
number, where a pair of in-gap helical edge states exists. Our findings
provide insights into the symmetry protected non-Hermitian topological
insulators.
\end{abstract}

\maketitle

\section{Introduction}

The topological phase of matter in condensed matter physics has been
attracting considerable research interest and has been widely explored \cite%
{Kane,Xiao,SCZhang,Ryu,Das,Wen,Vishwanath}. Open systems ubiquitously exist
in physics \cite{BenderRPP,NM,RotterRPP}, particularly, the optical and
photonic systems; these are mostly non-Hermitian because they interact with
the environment \cite{FL,EL,YFChen,Christodoulides,VKonotop,VSuchkov}.
Currently, topological systems extend into the non-Hermitian region \cite%
{Rudner,Szameit,Esaki,Diehl,Hu,GQLiang,Zeuner,Malzard,Rudner16,XWLuo18,HChenPRL2018,OZ,WPSR,WJGong,ZXZ19,Bergholtz1903,Kartashov15,CHe,Kunst2,WBRui,LXiao,Rakovszky,CYin,HJiang,YWang,SLPRA,KLZSR,Bergholtz1903,Yamamoto}%
, and the nontrivial topological properties are studied in one-dimensional
(1D), two-dimensional (2D), and three-dimensional (3D) systems, including
the Su-Schrieffer-Heeger (SSH) model \cite%
{SchomerusOL,YuceSSH,LJLSSH,Lieu,JLSR}, Aubry-Andr\'{e}-Harper (AAH) model
\cite{Joglekar,LJLAA,YXu1901,SLonghiPRL,YuceAA,Conti}, Rice-Mele (RM) model
\cite{WR,Kunst}, Chern insulator \cite%
{Philip,HZhai,ZWang2,Kawabata,Ezawa1904}, and Weyl semimetal \cite%
{YXu,CerjanPRB}.

Edge states in the parity-time ($\mathcal{PT}$)-symmetric systems are
considered absent because the edge state with its probability localized at
one system boundary is not $\mathcal{PT}$ invariant \cite{Hu}. However, $%
\mathcal{PT}$-symmetric interface states can localize at the interface or\
domain wall of two configurations with different topologies \cite%
{Poli,Weimann}; a zero mode can appear at the interface of lattices in the
same topological phase but with different non-Hermitian phases \cite{Pan}.
Both the types of interface states are topologically protected and robust to
disorders. Topological edge states lasing in 1D \cite%
{Amo,Khajavikhan,Feng,SLonghiLasing} and 2D \cite%
{Harari,Bandres,Kartashov,MSecli} have been reported recently; robust
single-mode lasing prevails due to the topological protection. Non-Hermitian
topological systems with $\mathcal{PT}$ symmetry are of primary focus \cite%
{LJin17,Menke,Ghatak,XNi,KawabataPT,Takata,Yoshida1904}. Other symmetries,
such as chiral-time ($\mathcal{CT}$) and charge-parity ($\mathcal{CP}$)
symmetries, have been investigated \cite%
{CZhang1904,SLPRA,LJLSSH,Malzard,Cancellieri}. The topological aspects of
non-Hermitian systems, including edge modes \cite{Borgnia,Cancellieri},
topological invariant \cite%
{GhatakReview,LiangHunagPRA,Ohashi,Leykam,YXu,LFu,FSong1905,SLin}, band
theory \cite{LFu,Papaj,Yokomizo,ZYGe}, topological pumping \cite%
{HZhai,WR,WRPRB,ChongPump,YucePump}, classification \cite%
{ZGong,Kawabata1812,KawabataNC,HZhou,Kawabata1902,LLi,ShuChenPRB},
high-order non-Hermitian topological systems \cite%
{TLiu,CHLeeHOT,Edvardsson,Ezawa,XWLuo1903,ZZhang}, semimetals \cite%
{Schmidt,Molina,Carllstrom,CHLeeTidal,Zyuzin,RAMolina}, and
symmetry-protected topological phases and localized states have been
investigated \cite{SLin,Budich,Okugawa,Yoshida,JunpengHou}. The exceptional
points (EPs) of non-Hermitian systems are connected by Fermi arcs \cite%
{LFu17,TYoshida,HZhouScience,HS} or form EP rings and surfaces \cite%
{BZhen,CerjanPRB,CerjanEPring,HZhouOptica}.

Notably, the bulk-boundary correspondence fails in some non-Hermitian
topological systems \cite%
{TELee,Xiong,Torres1,Torres2,ZGong,Kunst,ZWang1,ZWang2,JL18,PWang19,HZhang,CHLee,RYu,JHu,Herviou,WYi,FSong}%
. The spectrum under the periodical boundary condition (PBC) significantly
differs from that under the open boundary condition (OBC), and the
eigenstates under OBC are all localized at the system boundary (the
non-Hermitian skin effect) \cite{ZWang1}. The topological invariant can be
constructed either from the biorthogonal norm \cite{Kunst}, the non-Bloch
bulk \cite{ZWang1,ZWang2}, or the singular-value decomposition of the
Hamiltonian \cite{Herviou}. The reason for the breakdown of bulk-boundary
correspondence is that an asymmetric coupling induces an imaginary
Aharonov-Bohm effect \cite{JL18,PWang19}; the validity of the bulk-boundary
correspondence can be maintained by chiral-inversion symmetry \cite{JL18}.
The boundary modes in non-Hermitian systems have been discussed on the basis
of the transfer matrix method \cite{Kunst1812} and the Green's function
method \cite{Borgnia,Zirnstein,Silveirinha}. The interplay between
non-Hermiticity and non-Abelian gauge potential has been discussed \cite%
{JQCai}. In contrast to the topological phase transition, the invalidity of
bulk-boundary correspondence, and the non-Hermitian skin effect,
non-Hermiticity may not alter the topological phase transition and system
topology \cite{KLZhang}. A graphical approach has been proposed to visualize
the topological phases in non-Hermitian systems \cite{XMYang}.

Different from the non-Hermitian Chern insulators in Refs.~\cite%
{Philip,HZhai,ZWang2,Kawabata,Ezawa1904}, an inversion symmetric 2D
non-Hermitian Chern insulator is proposed and the validity of conventional
bulk-boundary correspondence is predicted in the end of Ref.~\cite{JL18}.
However, more physical aspects including the topological invariant and the
edge state are not discussed. In this study, the inversion symmetric
non-Hermitian Chern insulator is investigated in detail; in particular, the
topological properties of the gapless phase as well as the technical aspect
of topological characterization are systematically analyzed. Notably, the
gain and loss are alternately added in the $x$ and $y$ directions in the
inversion symmetric non-Hermitian Chern insulator. The non-Hermitian
Aharonov-Bohm effect and skin effect are prevented by the inversion
symmetry, and the conventional bulk-boundary correspondence is valid. The
Chern number constructed from the system bulk is a topological invariant
used to predict the topological properties of the Chern insulator. Different
from the anomalous edge states that are localized in a single unit cell \cite%
{TELee}, and those that cannot be predicted by the bulk topology \cite%
{Kawabata}, a pair of helical edge states appear in the topologically
nontrivial phase of the inversion symmetric Chern insulator under OBC. The
gapless phase has band touching points in the Brillouin zone (BZ). The
locations of the band touching degenerate points (DPs) are fixed and they do
not change into pairs of EPs \cite{HZhouScience}. The band touching EPs are
isolated and topologically protected, moving in the BZ and merging when they
meet in pairs. Band touching varies the system topology and the existence of
helical edge states. Moreover, non-Hermiticity creates a pair of
topologically protected in-gap helical edge states in a novel phase with
zero Chern number.

The remainder of the paper is organized as follows. In Sec.~\ref{II}, we
introduce the inversion symmetric 2D non-Hermitian Chern insulator. In Sec.~%
\ref{III}, we discuss the energy bands, the phase diagram, and the
topological characterization of the bulk Hamiltonian. In Sec.~\ref{IV}, we
demonstrate the energy spectrum and the helical edge states in different
topological phases of the edge Hamiltonian, and we verify the validity of
conventional bulk-boundary correspondence. In Sec.~\ref{V}, we present the
connection between the 2D non-Hermitian Chern insulator and other quasi-1D
non-Hermitian topological systems with asymmetric couplings. In addition, we
discuss possible experimental realization. Finally, in Sec.~\ref{VI}, we
summarize the main findings.

\section{Inversion symmetric non-Hermitian Chern insulator}

\label{II} 
\begin{figure}[tb]
\includegraphics[bb=0 0 310 215, width=8.7 cm, clip]{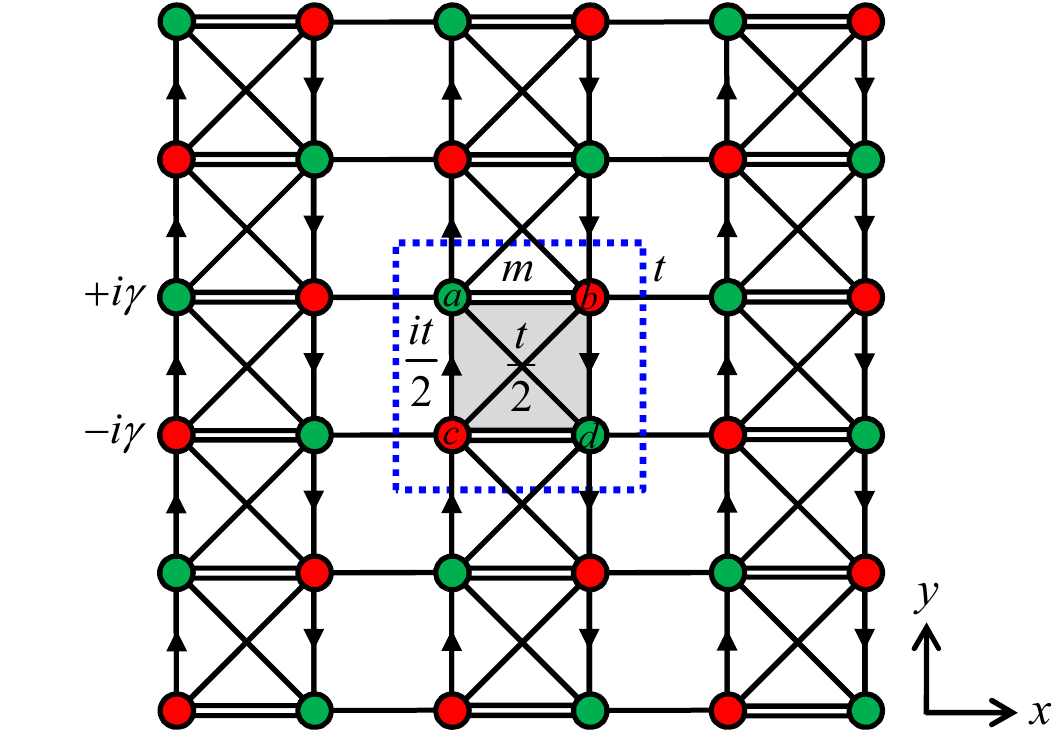}
\caption{Schematic of the inversion symmetric 2D non-Hermitian Chern
insulator. The green (red) solid circle indicates the site with gain (loss).
The Peierls phase factor is $e^{i\protect\pi/2}$ in the nonreciprocal
coupling $it/2$ in the vertical direction. The Peierls phase leads to a flux
of $\protect\pi$ in each plaquette.} \label{fig1}
\end{figure}

We investigate a non-Hermitian 2D topological system with
inversion symmetry. The schematic of the lattice of the non-Hermitian 2D
topological system is presented in Fig.~\ref{fig1}. The lattice in the
Hermitian case is a Chern insulator, constituted by vertical Creutz ladders
\cite{Creutz}. The Creutz ladders are horizontally coupled with strength $t$%
. The\ intra-ladder rung represents a coupling with strength $m$. Couplings
of strength $m$ and $t$ are alternated along the $x$ direction. Along the $y$
direction, the $t/2$ coupling\ has an additional $\pm \pi /2$ Peierls phase
factor in the front, which results in\ magnetic flux $\pi $ in each
plaquette, indicated by the shaded square in Fig.~\ref{fig1}. The
off-diagonal coupling strength in the plaquette is $t/2$, which is
equivalent to spin-orbital coupling \cite{SLZhuReview}. The lattice
Hamiltonian in the real space is $H=H_{0}+H_{1}$, consisting of the
nearest-neighbor coupling term%
\begin{eqnarray}
&&H_{0}=\sum_{i,j}[m(a_{i,j}^{\dagger }b_{i,j}+c_{i,j}^{\dagger
}d_{i,j})+t(a_{i+1,j}^{\dagger }b_{i,j}+c_{i+1,j}^{\dagger }d_{i,j})  \notag
\\
&&+\frac{it}{2}(a_{i,j}^{\dagger }c_{i,j}-a_{i,j-1}^{\dagger
}c_{i,j}-b_{i,j}^{\dagger }d_{i,j}+b_{i,j-1}^{\dagger }d_{i,j})]+\mathrm{%
H.c.,}
\end{eqnarray}%
and the off-diagonal coupling term%
\begin{equation}
H_{1}=\sum_{i,j}\frac{t}{2}(a_{i,j}^{\dagger }d_{i,j}+a_{i,j-1}^{\dagger
}d_{i,j}+b_{i,j}^{\dagger }c_{i,j}+b_{i,j-1}^{\dagger }c_{i,j})+\mathrm{H.c.,%
}
\end{equation}%
where the operators $a_{i,j}^{\dagger },b_{i,j}^{\dagger },c_{i,j}^{\dagger
},d_{i,j}^{\dagger }$ are creation operators for the four sublattices in the
unit cell $\left( i,j\right) $. The unit cell is indicated in Fig.~\ref{fig1}
by the blue square with dashed lines. The system exhibits inversion
symmetry, and the real space Hamiltonian $H$ is invariant under a rotation
of $\pi $ with respect to the lattice center. The inversion symmetry in 1D
systems is the reflection symmetry. The topological classification of
non-Hermitian systems with reflection symmetry and the rich topological
phases are studied \cite{ShuChenPRB}.

We consider generalization of the non-Hermitian system holding inversion
symmetry. The gain and loss with rates $\gamma $ are alternately introduced
in both $x$ and $y$ directions of $H$; the non-Hermitian Hamiltonian is $%
\mathcal{H}=H+H_{\gamma }$ and consists of staggered gain and loss%
\begin{equation}
H_{\gamma }=i\gamma \sum_{i,j}(a_{i,j}^{\dagger }a_{i,j}-b_{i,j}^{\dagger
}b_{i,j}-c_{i,j}^{\dagger }c_{i,j}+d_{i,j}^{\dagger }d_{i,j}).
\end{equation}%
Notably, the generalized non-Hermitian Chern insulator $\mathcal{H}$ holds
the inversion symmetry and differs from other non-Hermitian Chern insulators
without inversion symmetry in Ref.~\cite{Kawabata,ZWang2}. The
inversion symmetry ensures the presence of a zero imaginary magnetic flux
\cite{JL18}. Thus, the non-Hermitian skin effect is absent and the
conventional bulk-boundary correspondence holds in the inversion symmetric
non-Hermitian Chern insulator. The topological properties of the system can
be retrieved from its bulk topology. In the following sections, the energy
bands, phase diagram, topological characterization, and edge states are
elucidated.

\section{Phase diagram and topological characterization}

\label{III} \textit{Bloch Hamiltonian}. We apply the Fourier transformation $%
\rho _{k_{x},k_{y}}=N^{-1/2}\sum_{l,s}e^{-ik_{x}l}e^{-ik_{y}s}\rho _{l,s}$
to the sublattices $\rho =a,b,c,d$; the Hamiltonian of the non-Hermitian
system in the real space is rewritten in the momentum space, $\mathcal{H}%
=\sum_{\mathbf{k}}\mathcal{H}\left( \mathbf{k}\right) $. The Bloch
Hamiltonian of the non-Hermitian system in the basis $\{a_{k_{x},k_{y}}^{%
\dagger }|\mathrm{vac}\rangle ,b_{k_{x},k_{y}}^{\dagger }|\mathrm{vac}%
\rangle ,c_{k_{x},k_{y}}^{\dagger }|\mathrm{vac}\rangle
,d_{k_{x},k_{y}}^{\dagger }|\mathrm{vac}\rangle \}$ reads%
\begin{equation}
\mathcal{H}(\mathbf{k})=\left(
\begin{array}{cccc}
i\gamma & m+te^{-ik_{x}} & \Lambda _{-}(k_{y}) & \Lambda _{+}(k_{y}) \\
m+te^{ik_{x}} & -i\gamma & \Lambda _{+}(k_{y}) & -\Lambda _{-}(k_{y}) \\
\Lambda _{-}^{\ast }(k_{y}) & \Lambda _{+}^{\ast }(k_{y}) & -i\gamma &
m+te^{-ik_{x}} \\
\Lambda _{+}^{\ast }(k_{y}) & -\Lambda _{-}^{\ast }(k_{y}) & m+te^{ik_{x}} &
i\gamma%
\end{array}%
\right) ,
\end{equation}%
where we have $\Lambda _{+}\left( k_{y}\right) =t(1+e^{ik_{y}})/2$ and $%
\Lambda _{-}\left( k_{y}\right) =it(1-e^{ik_{y}})/2$. The Bloch Hamiltonian
has the inversion symmetry $\mathcal{P}\mathcal{H}\left( \mathbf{k}\right)
\mathcal{P}^{-1}=\mathcal{H}\left( -\mathbf{k}\right) $\ with $\mathcal{P}%
=\sigma _{x}\otimes \sigma _{x}$, and the particle-hole (charge-conjugation)
symmetry $\mathcal{CH}\left( \mathbf{k}\right) \mathcal{C}^{-1}=-\mathcal{H}%
\left( -\mathbf{k}\right) $\ with $\mathcal{C}=\left( \sigma _{0}\otimes
\sigma _{z}\right) \mathcal{K}$, where $\mathcal{K}$\ is the complex
conjugation operation; $\sigma _{x}$, $\sigma _{y}$, and $\sigma _{z}$ are
the Pauli matrices; and $\sigma _{0}$ is a $2\times 2$ identity matrix.

\begin{figure}[tb]
\includegraphics[ bb=0 0 420 610, width=8.7 cm, clip]{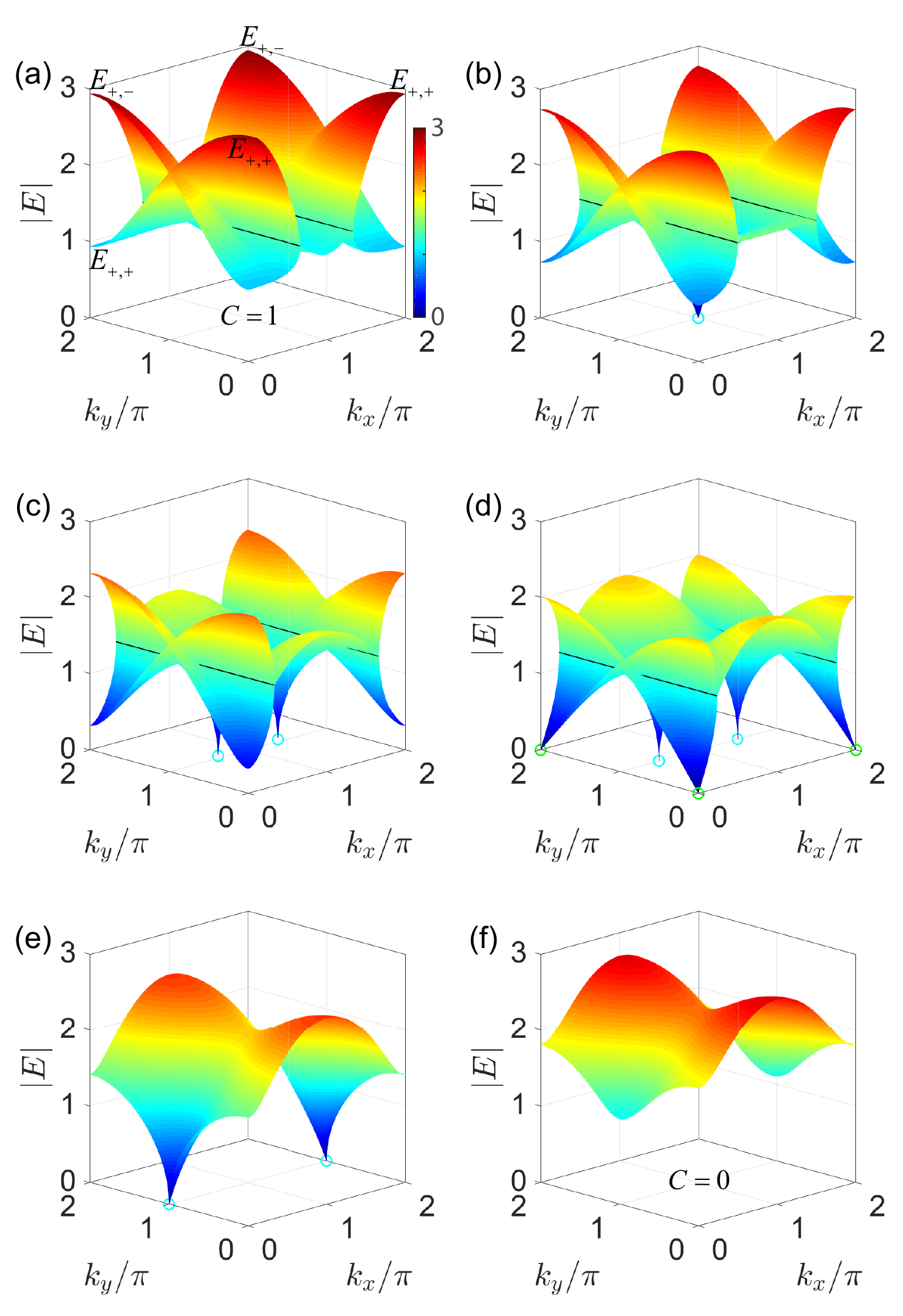}
\caption{Energy bands of the non-Hermitian Chern insulator $\mathcal{H}(\mathbf{k})$ at $m=t$ for various $\protect\gamma$. (a) $\protect\gamma=1/2$, (b) $\protect\gamma=1$, (c) $\protect\gamma=3/2$, (d) $\protect\gamma=\protect\sqrt{3}$, (e) $\protect\gamma=\protect\sqrt{5}$, and (f) $\protect\gamma=5/2$. The parameters are indicated by the hollow blue circles in the
phase diagram shown in Fig.~\protect\ref{fig3}. The absolute value of the
upper band energy $|E_{+,\pm }(\mathbf{k})|$ is depicted. The EPs (DPs) in
the gapless phase are marked by the cyan (green) circles. The black lines
indicate the energy at the HDELs. All color bars are identical to
the one shown in (a).} \label{fig2}
\end{figure}

The particle-hole symmetry ensures the spectrum of $\mathcal{H}\left(
\mathbf{k}\right) $ to be symmetric about zero energy, with the energy bands
given by%
\begin{equation}
E_{\pm ,\pm }(\mathbf{k})=\pm \sqrt{h_{x,\pm }^{2}+h_{y}^{2}+h_{z}^{2}}.
\label{E4}
\end{equation}%
where $\mu =m+t\cos k_{x}+\gamma $, $\nu =m+t\cos k_{x}-\gamma $, and $%
h_{x,\pm }=\sqrt{\mu \nu }\pm t\cos (k_{y}/2)$, $h_{y}=t\sin (k_{y}/2)$, $%
h_{z}=t\sin k_{x}$. For $\mu \nu <0$, $\sqrt{\mu \nu }$ is imaginary and the
energy spectrum is complex. Highly defective exceptional lines
(HDELs) appear at $\mu \nu =0$, that is, when $t\cos k_{x}=-m\pm \gamma $. The HDELs are fully constituted by EPs, and the energy levels
are two-state coalesced in pairs at energy $\pm \sqrt{t^{2}+h_{z}^{2}}$. The
energies are indicated by the black lines in Fig.~\ref{fig2}. Notably, the HDELs are EP lines across the BZ. The upper and lower bands
shrink into two levels with opposite energies. In the BZ, zero to four HDELs symmetrically appear about $k_{x}=0$ at different system
parameters. In addition, the bandgap closes at zero energy, where the band
touching points can be either DPs [$h_{x,-(+)}=h_{y}=h_{z}=0$] or EPs. The
inversion symmetry leads to an inversion symmetric distribution of the band
touching points in the BZ and the band touching points always appear in
pairs of $\left( k_{x},k_{y}\right) $ and $\left( -k_{x},-k_{y}\right) $.
Figure~\ref{fig2} depicts the absolute value of the energy spectrum $|E_{{+}%
,\pm }(\mathbf{k})|$. From Fig.~\ref{fig2}(a) to Fig.~\ref{fig2}(f), the
non-Hermiticity increases from $\gamma /t=1/2$ to $\gamma /t=5/2$. The
energy bands satisfy $E_{+(-),+}(k_{x},k_{y})=E_{+(-),-}(k_{x},k_{y}+2\pi )$%
. The two upper (lower) energy bands of the inversion symmetric
non-Hermitian Chern insulator constitute an entire band without intersection
if one of the two bands $E_{+,\pm }(\mathbf{k})$ [$E_{-,\pm }(\mathbf{k})$]
shifted by $2\pi $ along the $k_{y}$ direction in the BZ. The constituted
band is the upper (lower) band of Eq.~(\ref{hk}) by substituting $k_{y}$
with $k_{y}/2$; the corresponding energy bands are depicted in the Appendix
A.

\begin{figure}[tb]
\includegraphics[ bb=0 20 480 395, width=8.8 cm, clip]{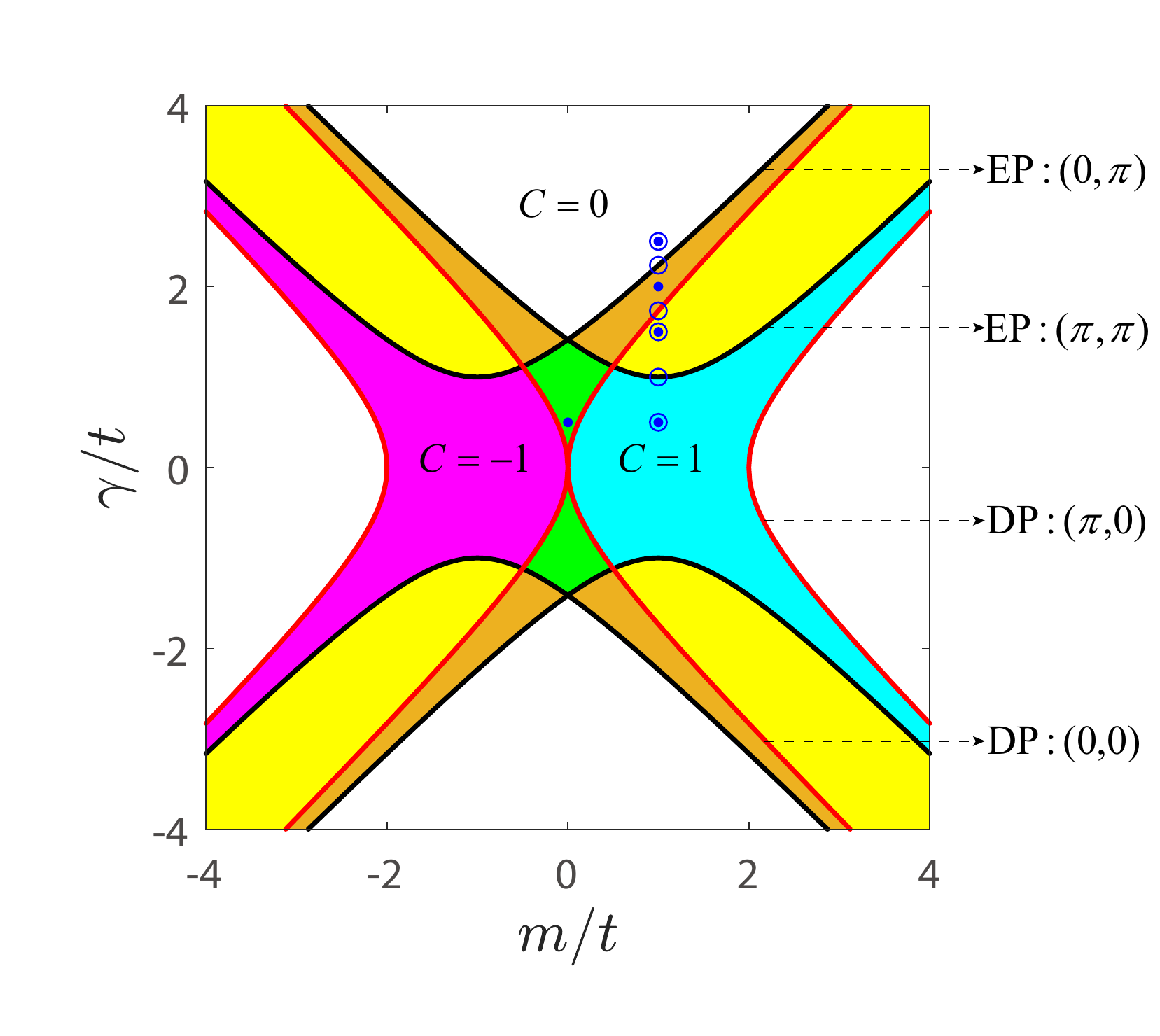}
\caption{Phase diagram in the $m$-$\protect\gamma$ parameter space. The
yellow and orange regions indicate the gapless phase. In the phase with
separable bands, the Chern number is nonzero in the magenta and cyan
regions; the Chern number is zero in the white and green regions. The black
(red) curves indicate the gapless phase with EPs (DPs), where the Chern
number is not available and winding number [Eq.~(\protect\ref{w})] is
employed as a topological invariant. For the topological phase with nonzero
Chern number, the helical edge states exist for the system under OBC in the $x$ direction or in the $y$ direction. In the gapless phase, the helical edge
states exist for the system under OBC only in one direction. In-gap helical
edge states exist in the green region. The hollow blue circles (solid blue
dots) indicate the parameters chosen in Fig.~\protect\ref{fig2} (Fig.~\protect\ref{fig6}).}
\label{fig3}
\end{figure}

\textit{Phase diagram. }To analyze the phase diagram of the system in
detail, we consider $t$ as the unit. The phase diagram of the non-Hermitian
Chern insulator is depicted in Fig.~\ref{fig3} as a function of system
parameters $m$ and $\gamma $. In the Hermitian situation with $\gamma =0$,
the energy bands are gapped for $m/t\neq 0,\pm 2$ and the Chern number is $%
C=1$ ($C=-1$) for $0<m/t<2$ ($-2<m/t<0$). In case $\left\vert m/t\right\vert
>2$, the Chern number is zero. In the non-Hermitian situation with $\gamma
\neq 0$, instead of splitting into pairs of EPs \cite{HZhouScience}, the
band touching DPs appear at fixed positions in the BZ, with their appearance
distinguishing the topologically nontrivial and trivial phases. In the
gapless phase, the band touching DPs occur at $\sqrt{\mu \nu }-t\cos
(k_{y}/2)=h_{y}=h_{z}=0$, which depends on non-Hermiticity given by%
\begin{equation}
\gamma ^{2}=\left( m\pm t\right) ^{2}-t^{2}.
\end{equation}%
This is indicated by the red curves in the phase diagram in Fig.~\ref{fig3}.
The DP locations are fixed in the BZ and at $\left( k_{x},k_{y}\right)
=\left( 0,0\right) $ for $\gamma ^{2}=\left( m+t\right) ^{2}-t^{2}$ and at $%
\left( k_{x},k_{y}\right) =\left( \pi ,0\right) $ for $\gamma ^{2}=\left(
m-t\right) ^{2}-t^{2}$.

The yellow and orange regions in Fig.~\ref{fig3} represent the gapless phase
with two symmetry protected EPs. The EPs possess distinct topology from that
of DPs due to the bifurcation linking the Riemann surface \cite%
{HeissEP,Dembowski,Berry,Uzdin,Doppler,Xu,Ding,CTChanPRX2018,Wiersig14,ZPLiu,WChenSensing,HodaeiSensing,Midya,AAlu}%
. The band touching points are EPs with fractional charge $\pm 1/2$\
according to the definition of non-Hermitian winding number $\mathcal{W}_{%
\mathrm{EP}}=-\left( 2\pi \right) ^{-1}\oint_{\Gamma }\nabla _{\mathbf{k}%
}\arg \left[ E_{+}\left( \mathbf{k}\right) -E_{-}\left( \mathbf{k}\right) %
\right] d\mathbf{k}$ \cite{LFu}, where $\Gamma $ is a closed loop in the
momentum space. In these gapless regions, the symmetry protected EPs move in
the BZ as system parameters, and merge when they meet in the BZ due to
topological phase transition; then, the number of EPs in the BZ reduces to
one and the topological features of EP change. As $\gamma $ increases, the
separate upper and lower bands become closer and may touch at $\left\vert
\gamma /t\right\vert >1$. The band touching EPs appear in the regions%
\begin{equation}
\left( \left\vert m\right\vert -t\right) ^{2}+t^{2}\leqslant \gamma
^{2}\leqslant \left( \left\vert m\right\vert +t\right) ^{2}+t^{2}.
\end{equation}%
At $m=0$, we have $\gamma ^{2}=2t^{2}$; the EPs form EP lines along $%
k_{y}=\pi $. Otherwise at $m\neq 0$, two EPs appear at $\left(
k_{x},k_{y}\right) =\left( \pm \theta ,\pi \right) $ when the system is
within the gapless regions, where $\theta $ is obtained from $\cos \theta
=\left( \gamma ^{2}-m^{2}-2t^{2}\right) /\left( 2mt\right) $. The EPs in the
BZ appear and move along $k_{y}=\pi $ as system parameters, and
symmetrically distribute about $k_{x}=0$ with different chiralities \cite%
{HeissEP,Dembowski}. At the boundary of the gapless phase represented by the
black curves in the phase diagram Fig. \ref{fig3}, two EPs merge to one hybrid EP~\cite{LFu,SLin,LJin1908} and locate at $\left(
k_{x},k_{y}\right) =\left( 0,\pi \right) $\ for $\gamma ^{2}=\left(
m+t\right) ^{2}+t^{2}$\ and at $\left( k_{x},k_{y}\right) =\left( \pi ,\pi
\right) $\ for $\gamma ^{2}=\left( m-t\right) ^{2}+t^{2}$; the
EP can be checked by substituting $(k_{x},k_{y})$ into $\mathcal{H}(\mathbf{k%
})$. The band structures in the gapless phase with different EP
distributions are demonstrated in Figs. \ref{fig2}(b)-\ref{fig2}(e).

The white region in the phase diagram (Fig. \ref{fig3}) indicate the
topologically trivial phase with zero Chern number; topologically protected
edge states are absent in this phase under OBC in either direction. Due to
the large imaginary part of energy bands at large non-Hermiticity of $\gamma
^{2}>\left( \left\vert m\right\vert +t\right) ^{2}+t^{2}$, the energy bands
are fully complex and separable. The green region also represent a phase
with zero Chern number; however, in-gap topologically protected edge states
exist under OBC in the $y$ direction; this point is elucidated in Sec.~\ref%
{IV}.

To characterize the topological properties of the inversion symmetric
non-Hermitian Chern insulator, it is convenient to transform the Bloch
Hamiltonian $\mathcal{H}\left( \mathbf{k}\right) $ into a two-band model.
After a similar transformation%
\begin{equation}
\mathcal{M}=\left(
\begin{array}{cccc}
\sqrt{\mu } & -i\sqrt{\mu } & 0 & 0 \\
-i\sqrt{\nu } & \sqrt{\nu } & 0 & 0 \\
0 & 0 & \sqrt{\nu } & -i\sqrt{\nu } \\
0 & 0 & -i\sqrt{\mu } & \sqrt{\mu }%
\end{array}%
\right) ,
\end{equation}
is performed, the Bloch Hamiltonian of the non-Hermitian Chern insulator
changes into%
\begin{equation}
\mathcal{MH}\left( \mathbf{k}\right) \mathcal{M}^{-1}=\left(
\begin{array}{cccc}
t\sin k_{x} & \sqrt{\mu \nu } & 0 & te^{ik_{y}} \\
\sqrt{\mu \nu } & -t\sin k_{x} & t & 0 \\
0 & t & t\sin k_{x} & \sqrt{\mu \nu } \\
te^{-ik_{y}} & 0 & \sqrt{\mu \nu } & -t\sin k_{x}%
\end{array}%
\right) ,  \label{MHM}
\end{equation}%
which possesses a two-site unit cell instead of a four-site one because of
the repeated diagonal $2\times 2$ matrix $\sqrt{\mu \nu }\sigma _{x}+\left(
t\sin k_{x}\right) \sigma _{z}$. This is in accordance with the spectrum
presented in Fig. \ref{fig2}, where two upper (lower) bands can form a
single band after one of them shifts by $2\pi $ in the $k_{y}$ direction.

The equivalent two-band Bloch Hamiltonian $h(\mathbf{k})=\mathbf{B}\cdot
\mathbf{\sigma }$ obtained from $\mathcal{H}(k)$ after a similar
transformation [Eq.~(\ref{MHM})] reads
\begin{equation}
h(\mathbf{k})=B_{x}\sigma _{x}+B_{y}\sigma _{y}+B_{z}\sigma _{z}.  \label{hk}
\end{equation}%
The Bloch Hamiltonian is a spin-1/2 in a complex effective magnetic field $%
\mathbf{B=}\left( B_{x},B_{y},B_{z}\right) $ with $B_{x}=\sqrt{\mu \nu }%
+t\cos k_{y}$, $B_{y}=-t\sin k_{y}$, $B_{z}=t\sin k_{x}$; and $\mathbf{%
\sigma =}\left( \sigma _{x},\sigma _{y},\sigma _{z}\right) $. In the
Hermitian limit $\gamma =0$, the effective magnetic field $B_{x}=m+t\cos
k_{x}+t\cos k_{y}$, $B_{y}=-t\sin k_{y}$, $B_{z}=t\sin k_{x}$ is real, and $%
h\left( \mathbf{k}\right) $ follows the Qi-Wu-Zhang model \cite{QWZ}.
Notably, $k_y$ in the two-band $h(\mathbf{k})$ corresponds to $2k_{y}$ in
the four-band $\mathcal{H}(\mathbf{k})$. The energy bands of $h(\mathbf{k})$
are depicted in the Appendix A as a comparison with the corresponding energy
bands of $\mathcal{H} \left( \mathbf{k}\right) $ depicted in Fig.~\ref{fig2}.

We emphasize that $\mu \nu =0$ is the HDELs of the four-band
Bloch Hamiltonian $\mathcal{H}(\mathbf{k})$. In contrast, the equivalent
two-band Bloch Hamiltonian $h(\mathbf{k})$ is Hermitian at $\mu \nu =0$;
thus, the HDELs are absent in $h(\mathbf{k})$. $%
\mathcal{H}(\mathbf{k})$ and $h(\mathbf{k})$ are related by the similar
transformation $\mathcal{M}$ at $\mu \nu \neq 0$; they are not related by
the similar transformation $\mathcal{M}$ at $\mu \nu =0$ ($\mathcal{M}$$%
^{-1} $ does not exits) although their energy bands are identical. In this
sense, the HDELs are removed and absent in $h(\mathbf{k})$. Notably, the
wave function singularity in $\mathcal{H}(\mathbf{k})$ does not appear at
the HDELs; thus, the equivalent two-band Bloch Hamiltonian $h(\mathbf{k})$
carries the topological properties of $\mathcal{H}(\mathbf{k})$.

\textit{Chern number in the gapped phase.} The Bloch Hamiltonian $h(\mathbf{k%
})$ describes an RM ladder with glider reflection symmetry \cite{SLZhang,CLi}
consisting of two coupled RM chains, and the inter-ladder leg coupling is $%
\sqrt{\mu \nu }$. In the regions that the energy bands are separated~\cite%
{LFu}, the Chern number for the energy band is a topological invariant that
characterizes the topological properties and the appearance of edge states
of the system under OBC.

The eigenvalue is $\varepsilon _{\pm }(\mathbf{k})=\pm B$, where $%
B=(B_{x}^{2}+B_{y}^{2}+B_{z}^{2})^{1/2}$; the corresponding eigenstate is $%
\left\vert \psi _{\pm }(\mathbf{k})\right\rangle =N_{\pm }^{-1}[B_{z}\pm
B,B_{x}+iB_{y}]^{T}$, where $N_{\pm }=(|B_{z}\pm B|^{2}+\left\vert
B_{x}+iB_{y}\right\vert ^{2})^{1/2}$ and the eigenstate $|\psi _{\pm }(%
\mathbf{k})\rangle $ satisfies $\langle \psi _{\pm }(\mathbf{k})|\psi _{\pm
}(\mathbf{k})\rangle =1$. The wave function singularity occurs at $B_{z}\pm
B=B_{x}+iB_{y}=0$ \cite{Ryu} (Notably, the wave function
singularity is not refer to as the EP, where eigenstate coalescence). The
Berry connection for the eigenstate is $A_{\pm }(\mathbf{k})=-i\left\langle
\phi _{\pm }(\mathbf{k})\right\vert \nabla _{\mathbf{k}}\left\vert \psi
_{\pm }(\mathbf{k})\right\rangle $ \cite{YXu,Kawabata,ZWang2}, and the Berry
curvature is $\Omega _{\pm }(\mathbf{k})=\nabla _{\mathbf{k}}\times A_{\pm }(%
\mathbf{k})$, where $\left\vert \phi _{\pm }(\mathbf{k})\right\rangle $\ is
the eigenstate of $h^{\dagger }(\mathbf{k})$ with corresponding energy bands
$\varepsilon _{\pm }^{\ast }(\mathbf{k})$. The eigenstates of $h(\mathbf{k})$
and $h^{\dagger }(\mathbf{k})$ constitute a biorthogonal basis. $|\psi _{\pm
}(\mathbf{k})\rangle $ and $\left\vert \phi _{\pm }(\mathbf{k})\right\rangle
$ are known as the right and left eigenstates, and they satisfy biorthogonal
condition $\langle \phi _{\pm }(\mathbf{k})|\psi _{\pm }(\mathbf{k}^{\prime
})\rangle =\delta _{\mathbf{kk}^{\prime }}$ \cite{Ali,LJin2011}. The
biorthogonal condition does not settle the normalization coefficients of the
left and right eigenstates; associated with the normalization of right
eigenstate $\langle \psi _{\pm }(\mathbf{k})|\psi _{\pm }(\mathbf{k})\rangle
=1$, the normalization coefficients of eigenstates are fixed. The Chern
number is defined as the integration of $\Omega _{\pm }(\mathbf{k})$ over
the whole first BZ
\begin{equation}
C_{\pm }=\frac{1}{2\pi }\iint_{\mathrm{BZ}}\mathrm{d}k_{x}\mathrm{d}%
k_{y}\Omega _{\pm }.
\end{equation}%
{The Chern number can be alternatively defined from other choice of Berry
curvature based on the right and left eigenstates $\Omega _{\pm }^{RL}(%
\mathbf{k})=-i\nabla _{\mathbf{k}}\times \lbrack \left\langle \psi _{\pm }(%
\mathbf{k})\right\vert \nabla _{\mathbf{k}}\left\vert \phi _{\pm }(\mathbf{k}%
)\right\rangle ]$, solely based on the right eigenstate $\Omega _{\pm }^{RR}(%
\mathbf{k})=-i\nabla _{\mathbf{k}}\times \lbrack \left\langle \psi _{\pm }(%
\mathbf{k})\right\vert \nabla _{\mathbf{k}}\left\vert \psi _{\pm }(\mathbf{k}%
)\right\rangle ]$, and solely based on the left eigenstate $\Omega _{\pm
}^{LL}(\mathbf{k})=-i\nabla _{\mathbf{k}}\times \lbrack \left\langle \phi
_{\pm }(\mathbf{k})\right\vert \nabla _{\mathbf{k}}\left\vert \phi _{\pm }(%
\mathbf{k})\right\rangle ]$. Notably, the four definitions of Chern numbers
are equivalent~\cite{LFu,KLZhang}.}

The Chern numbers for the upper and lower bands are opposite $C\equiv
C_{-}=-C_{+}$. The Chern number for either band is capable of characterizing
the topological properties of the corresponding phases. The Chern number
different between two bands directly reflects the number of edge modes at
the interface of two distinct bulks. At large non-Hermiticity, $\mu \nu <0$,
the Berry curvature $\Omega _{\pm }(\mathbf{k})$ is well-defined in the BZ.
The wave function singularity results in a nonzero Chern
number, which predicts the nontrivial topology and the existence of edge
states. For the wave function with singularity, two gauges are used to
describe the wave function. $\left\vert \psi _{\pm }^{\mathbf{II}}(\mathbf{k}%
)\right\rangle $ replaces $\left\vert \psi _{\pm }^{\mathbf{I}}(\mathbf{k}%
)\right\rangle $ in an area $D $ that encloses the singularity of $%
\left\vert \psi _{\pm }^{\mathbf{I}}(\mathbf{k})\right\rangle $, that is, $%
\left\vert \psi _{\pm }^{\mathbf{II}}(\mathbf{k})\right\rangle =\left\vert
\psi _{\pm }^{\mathbf{I}}(\mathbf{k})\right\rangle e^{i\varphi _{\pm }^{R}(%
\mathbf{k})},\left\vert \phi _{\pm }^{\mathbf{II}}(\mathbf{k})\right\rangle
=\left\vert \phi _{\pm }^{\mathbf{I}}(\mathbf{k})\right\rangle e^{i\varphi
_{\pm }^{L}(\mathbf{k})}$. The phase dependence between two gauges results
in a relation between two Berry connections $A_{\pm }^{\mathbf{II}}(\mathbf{k%
})=A_{\pm }^{\mathbf{I}}(\mathbf{k})+\nabla _{\mathbf{k}}\varphi _{\pm }^{R}(%
\mathbf{k})$. The Stokes' theorem indicates that the Chern number equals the
winding of the variation of $\varphi _{\pm }^{R}$ along the loop that
encloses the area $D$.

\begin{figure}[tb]
\includegraphics[ bb=0 0 450 390, width=8.7 cm, clip]{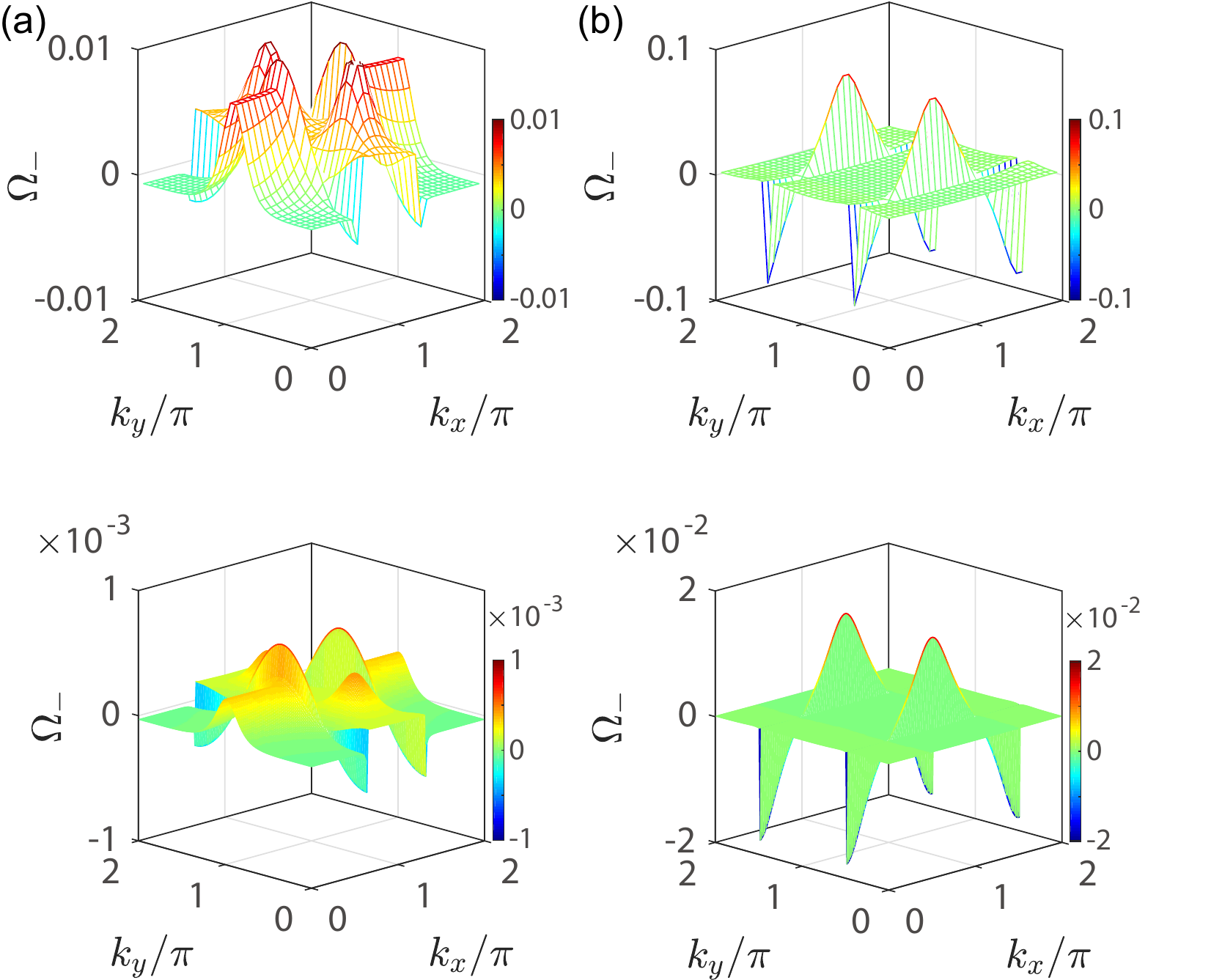}
\caption{Berry curvature in the BZ. Chern number as a summation of the Berry
curvature in the entire BZ yields $C=1.00$ for (a) $\protect\gamma=1/2$ and $C=0.00$ for (b) $\protect\gamma=5/2$ at $m=t=1$. The BZ is equally cut into $30\times30$ ($150\times150$) pieces in the numerical calculation for the
upper (lower) panel. At $k_x=2\protect\pi/3, 4\protect\pi/3$ in (a), $\protect\mu\protect\nu=0$ and $h(\mathbf{k})$ is Hermitian; $\protect\mu\protect\nu=0$ is absent in (b).}
\label{figChernN}
\end{figure}

The Berry curvature $\Omega _{\pm }(\mathbf{k})$ is ill-defined at $\mu \nu =0$, where eigenstates of $\mathcal{H}(\mathbf{k})$
coalesce in pairs and one-half of eigenstates vanish. Notably, $\sqrt{\mu
\nu }$ changes between real and imaginary as $k_{x}$ crosses $%
\mu \nu =0$ in the BZ. Consequently, $B_{x}$ ($h_{x,\pm }$) is not smooth
and the Berry curvature $\Omega _{\pm }(\mathbf{k})$ is ill-defined at $\mu \nu =0$. However, the Chern number is an
integral of the Berry curvature in the entire BZ, the ill-defined Berry
curvatures are only a finite number of lines; and the wave function
singularity does not appear at $\mu \nu =0$. Thus, the Chern number
determined by the wave function singularity in the BZ is not affected by the
presence of ill-defined Berry curvatures at $\mu \nu =0$. The nonzero
(zero) Chern number $C=\pm 1$ ($0$) is verified from the numerical
simulation in the discretized BZ \cite{Fukui}. The numerical results of the
Berry curvature are exemplified in Fig.~\ref{figChernN} for the
topologically nontrivial phase with $C=1$ and the topologically trivial
phase with $C=0$. In Fig.~\ref{figChernN}(a), $\mu \nu =0$ is the boundary
between Hermitian and non-Hermitian $h(\mathbf{k})$: $h(\mathbf{k})$ is
Hermitian when $\mu \nu >0$, but is non-Hermitian when $\mu \nu <0$.

\begin{figure}[tb]
\includegraphics[ bb=0 0 250 135, width=8.7 cm, clip]{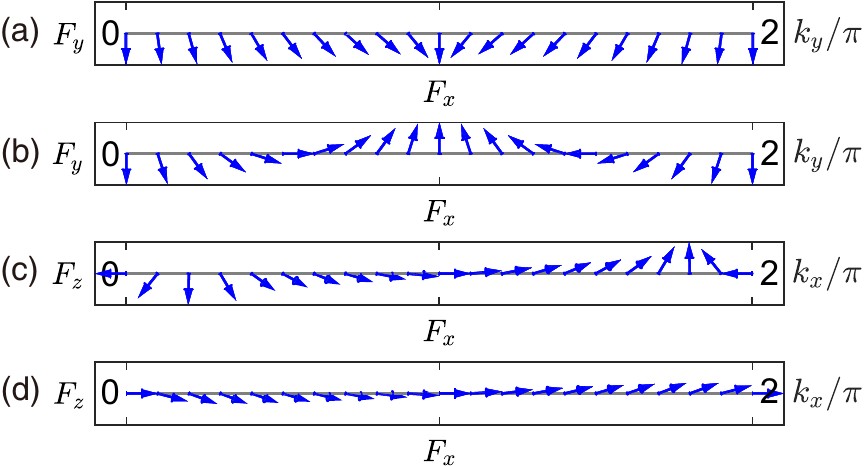}
\caption{The planar vector field $F_{-}$ for the lower band of $h(\mathbf{k}) $. (a) $\protect\gamma =3/2$, $k_{x}=0$; (b) $\protect\gamma =2$, $k_{x}=0$; (c) $\protect\gamma =3/2$, $k_{y}=\protect\pi $; (d) $\protect\gamma =2$, $k_{y}=\protect\pi $. $F_{-}$ at $k_y=\protect\pi$ in $h(\mathbf{k}) $
predicts the topological properties of $\mathcal{H}(\mathbf{k}) $ at $k_{y}=0 $. Other system parameters are $m=t=1$. (a) and (d) correspond to $w_{-}=0$; (b) and (c) correspond to $w_{-}=1$.}
\label{fig4}
\end{figure}

\textit{Winding number in the gapless phase.} In the gapless
phase, the Chern number is not available, we define a vector field $\mathbf{%
F}_{{\pm }}=( \left\langle \sigma_{x}\right\rangle _{{\pm }},\left\langle
\sigma _{y}\right\rangle_{{\pm } },\left\langle \sigma _{z}\right\rangle _{{%
\pm }}) $ to characterize the topological features of the gapless phase. The
vector field $\mathbf{F}_{\pm }$ is defined by the average values of Pauli
matrices: $F_{{{\pm},}x,y,z}=\left\langle \sigma _{x,y,z}\right\rangle _{{\pm%
}}=\left\langle \psi _{{\pm}}\left( \mathbf{k}\right) \right\vert\sigma
_{x,y,z}\left\vert \psi _{{\pm}}\left( \mathbf{k}\right)\right\rangle $ and
the subscript ${\pm}$ indicate the index of the energy band. Under the
normalization of the right eigenstate $\langle \psi _{{\pm}}(\mathbf{k}%
)|\psi _{{\pm}}(\mathbf{k})\rangle =1$, the amplitude of the vector field is
unity, that is, $\left\vert \mathbf{F}_{{\pm}}\right\vert ^{2}=1$. The
vector field is depicted in Fig.~\ref{fig4} to elucidate that a winding
number associated with $\mathbf{F}_{{\pm}}$ accurately predicts the
(non)existence of edge states in the gapless phase. For the system under OBC
in the $y$\ direction [Fig.~\ref{fig5}(b)], the trivial and nontrivial
windings of the planar vector field $\mathbf{F}_{{{\pm},}xy}=( \left\langle
\sigma _{x}\right\rangle _{{\pm} },\left\langle \sigma _{y}\right\rangle _{{%
\pm}}) $ as $k_{y}$ varying a period predict the attaching point of the edge
states and the bulk states in the complex energy bands
\begin{equation}
w_{{\pm}}=\left( 2\pi \right) ^{-1}\int_{0}^{2\pi }\mathrm{d}k_{y}\nabla
_{k_{y}}\phi _{{\pm}},  \label{w}
\end{equation}%
where $\tan \phi _{{\pm}}=F_{{{\pm},}y}/F_{{{\pm},}x}$. The winding numbers
of two energy bands are identical, that is, $w_{+}=w_{-}$. The nontrivial $%
2\pi $ varying direction accumulation of the planar vector field $\mathbf{F}%
_{{{\pm},}xz}=\left( \left\langle \sigma _{x}\right\rangle _{{\pm}
},\left\langle \sigma _{z}\right\rangle _{{\pm}}\right) $ in a period of $%
k_{x}$ predicts the edge states for the system under OBC in the $x $
direction [Fig.~\ref{fig5}(a)]; correspondingly, $w_{{\pm}}=\left( 2\pi
\right) ^{-1}\int_{0}^{2\pi }\mathrm{d}k_{x}\nabla _{k_{x}}\phi _{{\pm}}$
with $\tan \phi _{{\pm}}=F_{{{\pm},}z}/F_{{{\pm},}x}$ and $w_{{\pm}}$
predicts the topological phase transition at $k_{y}=\pi /2 $. Notably, the
topological phase transition at $k_{y}=\pi /2$ in $h\left( \mathbf{k}\right)
$ indicates the existence of edge states in the region $k_{y}=[-\pi ,\pi ]$
as depicted in Fig.~\ref{fig6} for $\mathcal{H}$ under OBC in the $x$
direction. This is because the folded BZ in the $k_{y}$ direction of $%
h\left( \mathbf{k}\right) $ yields the BZ of $\mathcal{H}\left( \mathbf{k}%
\right) $.

\section{Energy bands and Edge states of the Edge Hamiltonian}

\label{IV}

The bulk topology of the system determines the phase diagram, which
accurately predicts the topological phase transition and the (non)existence
of edge states in different topological phases. In this section, we
elucidate the role played by the Chern number for the gapped phase and the
winding number for the gapless phase in the topological characterization
through the investigation of the energy bands and edge states under
different OBC only in the $x$ and $y$ direction, respectively. The PBC and
OBC spectra are not dramatically different from each other owing to the
validity of conventional bulk-boundary correspondence protected by the
inversion symmetry.

\begin{figure}[b]
\includegraphics[ bb=0 0 280 155, width=8.8 cm, clip]{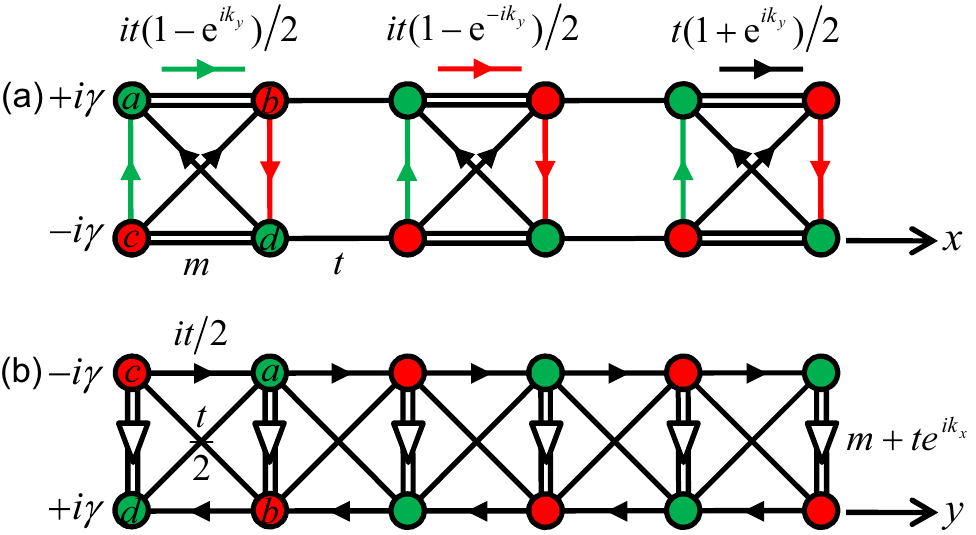}
\caption{(a) [(b)] Schematic of the 1D projection at set $k_y$ ($k_x$) of
the 2D inversion symmetric non-Hermitian Chern insulator. (a) Quasi-1D SSH
ladder. (b) Quasi-1D Creutz ladder. The green (red) solid circle indicates
the site with gain (loss).} \label{fig5}
\end{figure}

\begin{figure*}[tb]
\includegraphics[ bb=0 0 430 170, width=18 cm, clip]{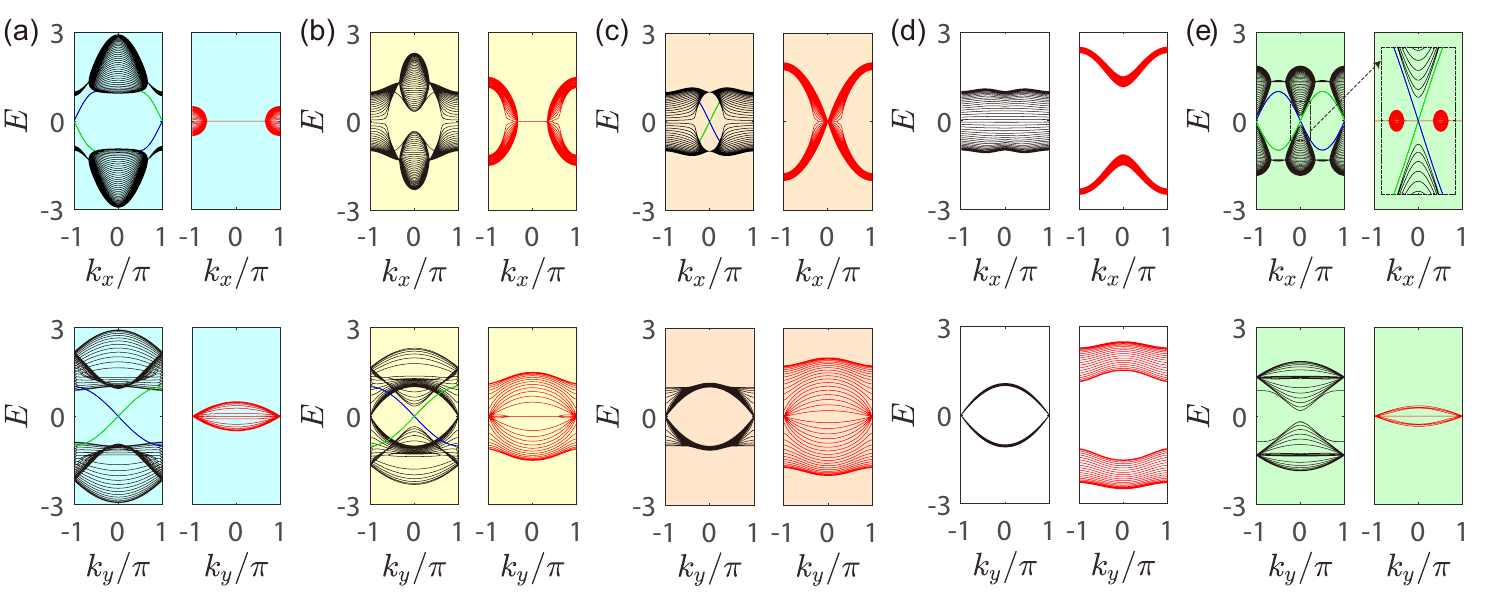}
\caption{Energy spectrum for $\mathcal{H}$ under OBC. The upper
(lower) panel represents the energy spectra of the system under PBC in the $%
x $ ($y$) direction, but under OBC in the $y$ ($x$) direction;
the schematic is in Fig.~\protect\ref{fig5}. The parameters are (a) $%
\protect\gamma =1/2$, (b) $\protect\gamma =3/2$, (c) $\protect\gamma =2$,
(d) $\protect\gamma =5/2$; other parameters are $m=t=1$ in (a)-(d). In (e), $%
m=0$, $t=1$, and $\protect\gamma =1/2$. The band energy repetition period is
$\protect\pi$ for the upper panel of (e), where the inset on the right side
are zoom-in plot of the area in the dashed box. The edge states reside
inside the bandgap without touching the bulk bands. The parameters in the
plots are indicated by the solid dots in the phase diagram shown in Fig.~%
\protect\ref{fig3}. The black (red) line is the real (imaginary) part of
band energy. The number of unit cells of 1D projection lattices is twenty.}
\label{fig6}
\end{figure*}
{\textit{1D edge Hamiltonian under OBC}.} The system under OBC is referred
to the edge Hamiltonian because it is generated by truncating the bulk
Hamiltonian in a certain way. The two-dimensional Chern insulator reduces to
a non-Hermitian quasi-1D SSH ladder or quasi-1D Creutz ladder at a fixed
momentum in the $y$ or $x$ direction (Fig.~\ref{fig5}), respectively. The
non-Hermitian quasi-1D Creutz ladder is equivalent to a 1D RM chain with
asymmetric couplings presented alternately (see Sec.~\ref{V} for more
detail).

For the Chern insulator under PBC in the $y$ direction and OBC in the $x$
direction, we apply the Fourier transformation in the $y$ direction. This
projection gives a quasi-1D SSH ladder with momentum $k_{y}$-dependent
couplings and staggered on-site potentials, as presented in Fig.~\ref{fig5}%
(a). Similarly, for the Chern insulator under PBC in the $x$ direction and
OBC in the $y$ direction, the projection lattice reduces to a quasi-1D
Creutz ladder with a momentum $k_{x}$-dependent couplings in the inter
ladder, as presented in Fig.~\ref{fig5}(b).

We study the inversion symmetric non-Hermitian Chern insulator under PBC and
OBC in the $y$ and $x$ directions, respectively. The upper panel of Fig.~\ref%
{fig6} depicts the spectra under PBC in the $x$ direction and under OBC in
the $y$ direction; the lower panel of Fig.~\ref{fig6} depicts the spectra
under PBC in the $y$ direction and under OBC in the $x$ direction at the
same system parameters as those in the upper panel of Fig.~\ref{fig6}.

{\textit{Gapped phase}.} In the gapped phase, the Chern number is the
topological invariant. The nonzero Chern numbers of the upper and lower
bands in the topologically nontrivial phase indicate the existence of a pair
of topologically protected helical edge states under OBC. The two helical
edge states in a pair localize at the left and right edges of the 1D system,
respectively. The white region with zero Chern number are topologically
trivial phases without topologically protected edge states. The green region
with zero Chern number consists of in-gap edge states. We first discuss the
topological edge states and then discuss the topological phase transition in
different gapped phases.

Figure~\ref{fig6}(a) depicts the spectrum in the topologically nontrivial
phase with $C=1$; a pair of topologically protected helical edge states
exist, localizing on the left and right boundaries of the 1D lattice,
respectively. In Fig.~\ref{fig6}(d), the system is in the topologically
trivial region, the white region with $C=0$, the edge state is absent. The
OBC spectra in the green region with $C=0$ are depicted in Fig.~\ref{fig6}%
(e). In-gap edge states are observed under OBC in the $y$ direction; both
edge states are detached from the upper and lower bands; this is indicated
by the zero Chern number and presented in the inset of Fig.~\ref{fig6}(e).
Two in-gap edge states cross at $k_{x}=0,\pi $. As non-Hermiticity
increases, the in-gap edge states may touch the upper and lower bands at $%
\gamma \geqslant 1$ with the diminishing gap at $k_{x}=\pm \pi /2$. By
contrast, the topologically protected helical edge states are absent for the
system under OBC in the $x$ direction in the lower panel. We generalize the
polarization into the non-Hermitian region to characterize the topological
in-gap edge states~\cite{FLiu,Resta}. The polarization is defined by the
Berry connection based on the left and right eigenstates; the projection in
the $x$ ($y$) direction is $P_{\pm ,x(y)}=\left( 2\pi \right) ^{-2}\int
\int_{\mathrm{BZ}}\mathrm{d}k_{x}\mathrm{d}k_{y}A_{\pm ,x(y)}\left( \mathbf{k%
}\right) $, where $A_{\pm ,x(y)}\left( \mathbf{k}\right) =-i\left\langle
\phi _{\pm }(\mathbf{k})\right\vert \nabla _{k_{x(y)}}\left\vert \psi _{\pm
}(\mathbf{k})\right\rangle $. In the green region with $C=0$, the wave
function has no singularities; thus, the polarization is well defined and we
have the 2D Zak phases associated fractional polarization $P_{-,y}=-{1/2}$
and $P_{-,x}=0$~\cite{FLiu}. The in-gap edge states are topological
protected, being robust to the inversion symmetric perturbations that
without breaking the pseudo-Hermiticity. Two in-gap edge states remain in
the bandgap and are detached from the bulk bands. The details about the
robustness of in-gap edge states are provided in the Appendix B.

The appearance of band touching affects the (non)existence of topologically
protected helical edge states. We elaborate this point by considering gapped
phases in the regions of $m>0$ and $\gamma <1$. Notably, the inversion
symmetric 2D non-Hermitian Chern insulator is anisotropic. In the green
region of the phase diagram, the helical edge states appear only in the
system under OBC in the $y$ direction {[upper panel of Fig.~\ref{fig6}(e)]}.
Between the green and cyan regions, the system experiences a band touching
with DP at $\left( k_{x},k_{y}\right) =\left( 0,0\right) $ [Fig.~\ref{fig2}%
(d)]. Thus, for the system under OBC in the $y$ direction, the in-gap
helical edge states crossing at $k_{x}=0,\pi $ change into helical edge
states that connect the upper and lower bands and cross only at $k_{x}=\pi $
{[upper panel of Fig.~\ref{fig6}(a)]}. As $m$ continuously increases, the
system experiences a band touching with DP at $\left( k_{x},k_{y}\right)
=\left( \pi ,0\right) $ and the helical edge states in the cyan region
crossing at $k_{x}=\pi $ are destroyed and disappear in the white region.
Similarly, for the system under OBC in the $x$ direction, the helical edge
states crossing at $k_{y}=0$ are formed after the system comes across the
band touching with DP at $\left( k_{x},k_{y}\right) =\left( 0,0\right) $ and
enters the cyan region in the phase diagram. As $m$ continuously increases,
the system experiences a band touching with DP at $\left( k_{x},k_{y}\right)
=\left( \pi ,0\right) $ and the helical edge states in the cyan region
crossing at $k_{y}=0$ are destroyed and disappear in the white region.

{\textit{Gapless phase}.} In the gapless phase, the helical edge states
appear only under OBC in one direction and are predicted by the winding
number $w_{\pm }$ associated with the vector field $\mathbf{F}_{\pm }$. The
nonzero winding of the vector field predicts the presence of edge states
detaching the bulk band. The planar vector field $\mathbf{F}_{-,xy}$ is
depicted in Figs.~\ref{fig4}(a) and~\ref{fig4}(b). The winding number is $%
w_{-}=0$ for Fig.~\ref{fig4}(a) and is $w_{-}=1$ for Fig.~\ref{fig4}(b),
which predict the absence and presence of edge states at $k_{x}=0$ for the
system under OBC in the $y$ direction in the upper panel of Figs.~\ref{fig6}%
(b) and~\ref{fig6}(c), respectively. The planar vector field $\mathbf{F}%
_{-,xz}$ in Figs.~\ref{fig4}(c) and~\ref{fig4}(d) are depicted. The winding
number is $w_{-}=1$ for Fig.~\ref{fig4}(c) and is $w_{-}=0$ for Fig.~\ref%
{fig4}(d), which predict the presence and absence of edge states at $k_{y}=0$
for the system under OBC in the $x$ direction in the lower panel of Figs.~%
\ref{fig6}(b) and~\ref{fig6}(c), respectively.

Topologically protected helical edge states exist in the orange region of
the phase diagram in the gapless phase only for the OBC in the $y$
direction. We consider a process with increasing non-Hermiticity. The system
first experiences a nontrivial phase with $C=1$ in the cyan region, then it
enters the gapless phase, and finally it stays in the trivial phase with $%
C=0 $ in the white region. For the system under OBC in the $y$ direction,
the helical edge states cross at $k_{x}=\pi $. As non-Hermiticity increases,
the system enters the gapless phase in the yellow region from the cyan
region. The EPs appear at $\left( k_{x},k_{y}\right) =\left( \pi ,\pi
\right) $ and then\ the $k_{x}$ position of EPs changes as $\gamma $
increases. Thus the helical edge states crossing at $k_{x}=\pi $ are
destroyed by non-Hermiticity [upper panel of Fig.~\ref{fig6}(b)]; the
crossing at $k_{x}=\pi $ disappear. However, the DP appears at $\left(
k_{x},k_{y}\right) =\left( 0,0\right) $ at increasing $\gamma $; then the
helical edge states cross at $k_{x}=0$ and reappear until the EPs move to $%
\left( k_{x},k_{y}\right) =\left( 0,\pi \right) $, which destroys the
helical edge states once again. Thereafter, the system enters the trivial
phase with $C=0$ in the white region. For instance, as non-Hermiticity
increases to $\gamma =\sqrt{3}$ for $m=t=1$, an additional band touching DP
is formed at $k_{x}=0$. Therefore, two helical edge states reappear in the
orange region of the phase diagram at large non-Hermiticity as shown in the
upper panel of Fig.~\ref{fig6}(c); they cross at $k_{x}=0$ instead of at $%
k_{x}=\pi $ as in the $C=1$ [upper panel of Fig.~\ref{fig6}(a)]. At even
larger $\gamma =5/2$ in Fig.~\ref{fig6}(d), the system is in the
topologically trivial region without the presence of edge states.

By contrast, the topologically protected helical edge states exist in the
yellow region of the phase diagram in the gapless phase only for the OBC in
the $x$ direction as elucidated in the lower panel of Fig. \ref{fig6}. For
the system under OBC in the $x$ direction, the helical edge states cross at $%
k_{y}=0$ [lower panel of Fig.~\ref{fig6}(a)]. The band touching EPs are
fixed at $k_{y}=\pi $ independent of non-Hermiticity, and do not affect the
helical edge states until the DP appears at $\left( k_{x},k_{y}\right)
=\left( 0,0\right) $ [Fig.~\ref{fig2}(d)]. Then, the helical edges crossing
at $k_{y}=0$ are destroyed and disappear in the gapless phase in the orange
region [lower panel of Fig.~\ref{fig6}(c)]. The gapless phase with band
touching DP is the boundary for the appearance of gapless topologically
protected edge states.

\begin{figure}[tb]
\includegraphics[ bb=0 0 560 530, width=7.5 cm, clip]{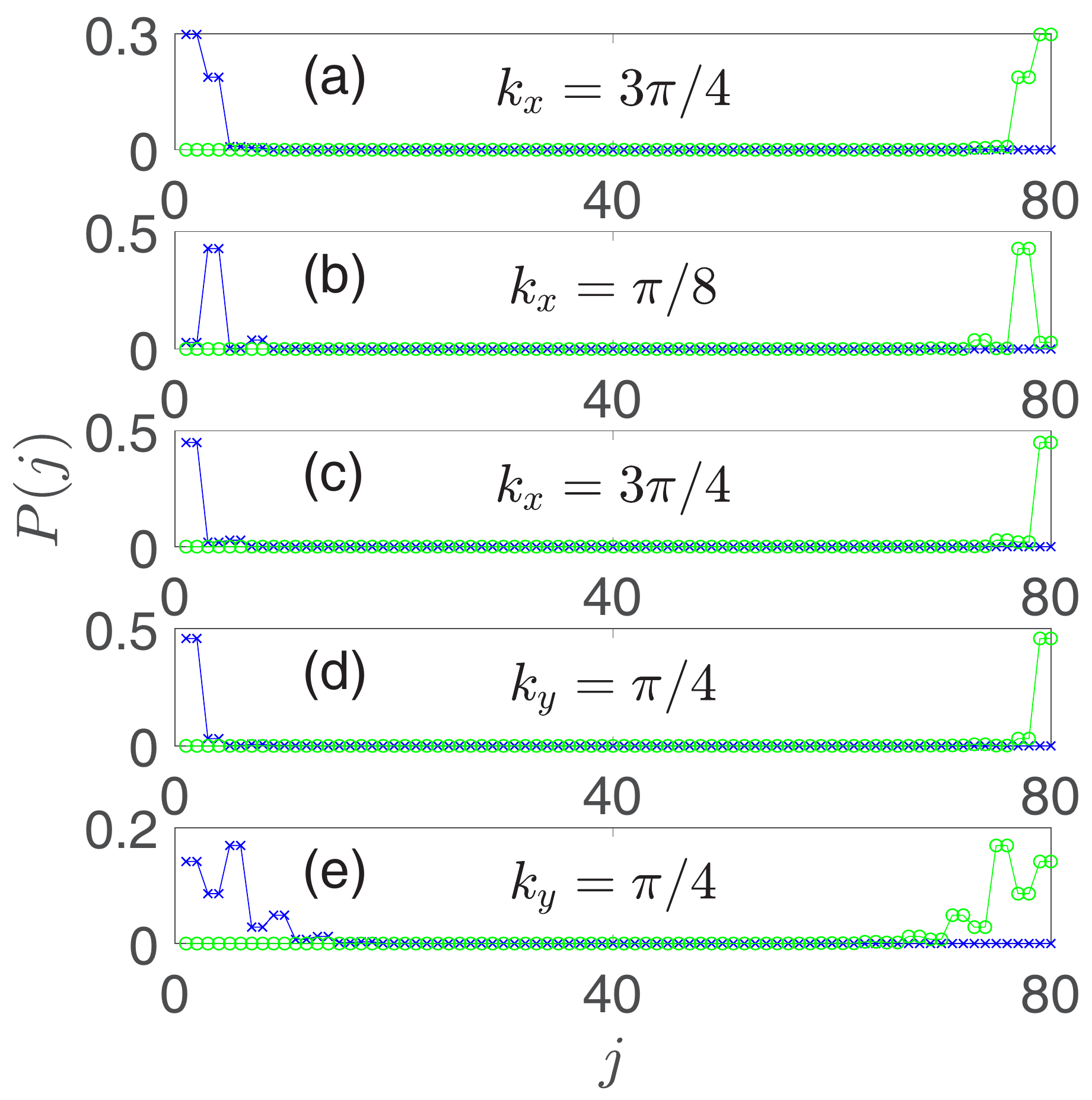}
\caption{Edge states for the quasi-1D lattice in Fig.~\ref{fig5} under OBC. The set
momentum in the direction under PBC is marked inside. The parameters are
identical with those chosen in Fig.~\protect\ref{fig6}. $\protect\gamma =1/2$
in (a), (c), (d); $\protect\gamma =2$ in (b); and $\protect\gamma =3/2$ in
(e). Other parameters are $m=t=1$ except for $m=0,t=1$ in (c). } \label{fig7}
\end{figure}

\textit{Edge states. }For the system under OBC in the $y$ direction, the
left edge state localizes at the left boundary and has eigen energy $%
E_{L}\left( k_{x}\right) =-t\sin k_{x}$. We consider $\psi _{1}=1$ for
convenience without loss of generality. The wave functions of the left edge
state $\left\vert \psi _{L}\right\rangle $ satisfies the recursion relation%
\begin{equation}
\left\{
\begin{array}{l}
\psi _{2j}=-i\psi _{2j-1}\text{,} \\
\psi _{2j+1}=[(-1)^{j}\gamma /t-\left( m+t\cos k_{x}\right) /t]\psi _{2j-1}%
\text{,}%
\end{array}%
\right.
\end{equation}%
where $j=1,2,3,\cdots ,n/2-1$ is the index, and $\psi _{n}=-i\psi _{n-1}$.
The right edge state $\left\vert \psi _{R}\right\rangle $, localizes at the
right boundary and has eigen energy $E_{R}\left( k_{x}\right) =t\sin k_{x}$.
We can consider $\psi _{n}=1$, and the recursion relation for the wave
functions of the right edge state $\left\vert \psi _{R}\right\rangle $ is%
\begin{equation}
\left\{
\begin{array}{l}
\psi _{n-(2j-1)}=-i\psi _{n-(2j-2)}\text{,} \\
\psi _{n-2j}=[(-1)^{j}\gamma /t-\left( m+t\cos k_{x}\right) /t]\psi
_{n-(2j-2)}\text{,}%
\end{array}%
\right.
\end{equation}%
where $j=1,2,3,\cdots ,n/2-1$ is the index, and $\psi _{1}=-i\psi _{2}$.
Notably, the edge state energy is independent of the coupling $m$ and the
non-Hermiticity $\gamma $ although both affect the wave functions. The edge
states under OBC in the $y$ direction in different topological phases are
depicted in Figs.~\ref{fig7}(a)-\ref{fig7}(c). Figure~\ref{fig7}(a) presents
the edge states in the $C=1$ phase, as illustrated in Fig.~\ref{fig6}(a);
Fig.~\ref{fig7}(b) represents the edge states in the gapless phase, as
illustrated in Fig.~\ref{fig6}(c); Fig.~\ref{fig7}(c) represents the in-gap
edge states in the $C=0$ phase, as illustrated in Fig.~\ref{fig6}(e). The
green region ($C=0$) with in-gap edge states is a novel phase induced by
non-Hermiticity.

For the system under OBC in the $x$ direction, the left edge state localizes
at the left boundary and has eigen energy $E_{L}\left( k_{y}\right) =-t\sin
\left( k_{y}/2\right) $. The wave functions of the left edge state $%
\left\vert \psi _{L}\right\rangle $ satisfies the recursion relation%
\begin{equation}
\left\{
\begin{array}{l}
\psi _{4j+1}=\frac{i\gamma }{t}\psi _{4j-1}-[\frac{m}{t}-\cos (\frac{k_{y}}{2%
})]\psi _{4j-3}, \\
\psi _{4j+3}=-\frac{\left( i\gamma /t\right) \psi _{4j+1}}{m/t+\cos (k_{y}/2)%
}-\frac{\psi _{4j-1}}{m/t+\cos (k_{y}/2)}, \\
\psi _{4j+2}=\psi _{4j+1}e^{i\theta }, \\
\psi _{4j+4}=-\psi _{4j+3}e^{i\theta },%
\end{array}%
\right.
\end{equation}%
where $j=1,2,3,\cdots ,n/4-1$ is the index, and $\psi _{1}=1$, $\psi
_{2}=e^{i\theta }$, $\psi _{3}=-\left( i\gamma /t\right) /[m/t+\cos
(k_{y}/2)]$, and $\psi _{4}=-\psi _{3}e^{i\theta }$ with $\sin \theta =\sin
(k_{y}/2)$, and $\cos \theta =-\cos (k_{y}/2)$. The right edge state,
localizes at the right boundary and has eigen energy $E_{R}\left(
k_{y}\right) =t\sin \left( k_{y}/2\right) $. The recursion relation for the
wave functions of the right edge state $\left\vert \psi _{R}\right\rangle $
is%
\begin{equation}
\left\{
\begin{array}{l}
\psi _{n-4j}=\frac{i\gamma }{t}\psi _{n-(4j-2)}-[\frac{m}{t}-\cos (\frac{%
k_{y}}{2})]\psi _{n-(4j-4)}, \\
\psi _{n-(4j+2)}=-\frac{\left( i\gamma /t\right) \psi _{n-4j}}{m/t+\cos
(k_{y}/2)}-\frac{\psi _{n-(4j-2)}}{m/t+\cos (k_{y}/2)}, \\
\psi _{n-(4j+1)}=\psi _{n-4j}e^{i\theta }, \\
\psi _{n-(4j+3)}=-\psi _{n-(4j+2)}e^{i\theta },%
\end{array}%
\right.
\end{equation}%
where $j=1,2,3,\cdots ,n/4-1$ is the index, and $\psi _{n}=1$, $\psi
_{n-1}=e^{i\theta }$, $\psi _{n-2}=-\left( i\gamma /t\right) /[m/t+\cos
(k_{y}/2)]$, and $\psi _{n-3}=-\psi _{n-2}e^{i\theta }$ with $\sin \theta
=\sin (k_{y}/2)$, and $\cos \theta =-\cos (k_{y}/2)$. The edge states under
OBC in the $x$ direction in different topological phases are depicted in
Figs.~\ref{fig7}(d) and \ref{fig7}(e). Figure~\ref{fig7}(d) represents the
edge states in the $C=1$ phase, as illustrated in Fig.~\ref{fig6}(a); Fig.~%
\ref{fig7}(e) represents the edge states in the gapless phase, as
illustrated in Fig.~\ref{fig6}(b).

\section{Discussion}

\label{V}

\begin{figure}[tb]
\includegraphics[ bb=0 0 290 160, width=8.7 cm, clip]{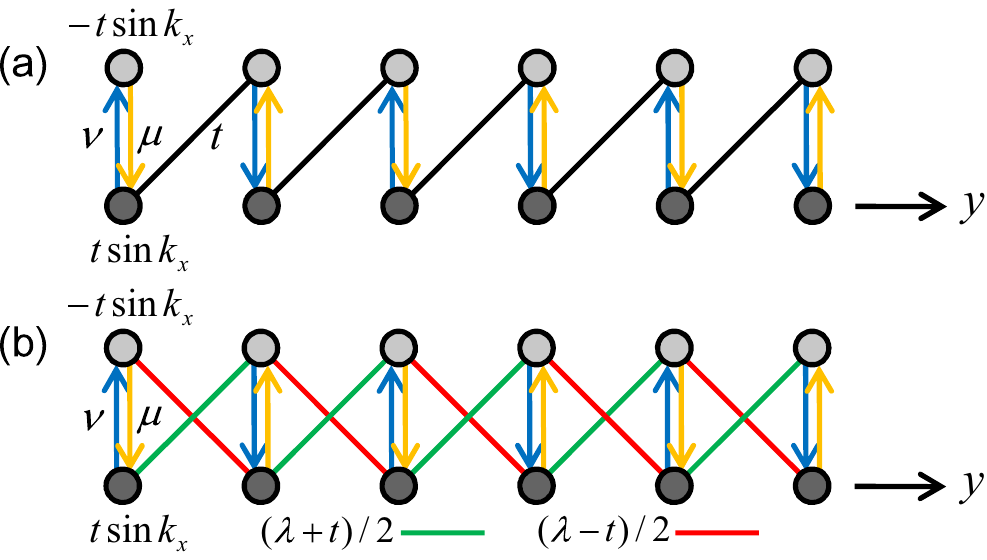}
\caption{(a) 1D RM chain. (b) Quasi-1D RM ladder with asymmetric
intra-ladder leg coupling at $\protect\lambda \neq \pm t$. (b) can be
regarded as a 1D RM chain of (a) with symmetric long-range coupling. $-t
\sin k_x$ and $t \sin k_x$ are the on-site real potentials for the upper
(light gray) and lower (dark gray) sites, respectively. The non-Hermitian
element is the asymmetric coupling $\protect\mu$-$\protect\nu$.} \label{fig8}
\end{figure}

\textit{1D projection of the Chern insulator.} The quasi-1D Creutz ladder
[Fig.~\ref{fig5}(b)] changes into a 1D RM\ ladder with asymmetric couplings
[Fig.~\ref{fig8}(a)] after applying a uniform transformation to each dimer
(the upper and lower sites with intra-dimer coupling $m$), $I_{2n}\otimes
\left( \sigma _{0}+i\sigma _{x}\right) /\sqrt{2}$. Moreover, if the
off-diagonal coupling strength in the plaquette is $\lambda /2$ instead of $%
t/2$ in Fig.~\ref{fig1}, we obtain a quasi-1D RM ladder presented in Fig.~%
\ref{fig8}(b), which can be regarded as a 1D RM chain with long-range
coupling or as two coupled RM chains with asymmetric inter-chain coupling.
The coupling depicted in green (red) is the nearest-neighbor coupling
between asymmetric dimers; the red (green) coupling is taken as the
long-range coupling, studied in Ref.~\cite{ZWang1}. In the situation that $%
\lambda =t$ $(\lambda =-t)$, the long-range couplings vanish. In the
neighboring asymmetric dimers, the asymmetric couplings with stronger and
weaker coupling amplitudes are in the opposite directions, which is\ due to
the difference in the one-way amplification and one-way attenuation. At $\mu
\nu =0$, the asymmetric couplings are unidirectional \cite{Longhi15,Midya}
and result in the HDELs. Notably, alternately introducing the
gain and loss under inversion symmetry prevents the one-way amplification,
one-way attenuation, and nonzero accumulation of imaginary flux. This is the
key point for the validity of conventional bulk-boundary correspondence.

\textit{Experimental realization.} The non-Hermitian Chern insulator can be
simulated by dissipative ultracold atoms in an optical lattice with a
synthetic magnetic field and the spin-orbital coupling \cite%
{Aidelsburger,Goldman,Cooper}. In addition, the non-Hermitian Chern
insulator can be implemented in optical and photonic systems such as coupled
waveguide and coupled resonator lattices \cite{Hafezi,Mittal}, where optical
dissipation and radiation ubiquitously exist. Instead of incorporating a
balanced gain and loss, introducing different dissipations in different
sublattices facilitates realization of passive non-Hermitian topological
systems in experiment. It is convenient to induce losses by sticking
additional absorption materials. In Ref.~\cite{TELee}, proposed realization
of a coupled resonator optical waveguide lattice of the non-Hermitian Creutz
ladder with gain and loss; this proposal is directly applicable realizing
the inversion symmetric non-Hermitian Chern insulator investigated in this
study by coupling the Creutz ladders together and adding the non-Hermiticity
alternately in the $x$ and $y$ directions.

\section{Summary}

\label{VI} In summary, we investigated an inversion symmetric 2D
non-Hermitian Chern insulator with balanced gain and loss in $x$ and $y$
directions and found that the conventional bulk-boundary correspondence
holds. The bulk topology determines the phase diagram and accurately
predicts the topological phase transition and the (non)existence of
topological edge states for the system under OBC. The helical edge states
exist in the phase with nonzero Chern number for the system under OBC in
both directions. By contrast, non-Hermiticity can vary the system topology
and destroy (create) helical edge states. Non-Hermiticity creates
topological gapless phase, where the helical edge states exist in the
inversion symmetric non-Hermitian Chern insulator under OBC in only one
direction. The winding number associated with the vector field of the
average values of Pauli matrices predicts the edge states in the gapless
phase. Furthermore, non-Hermiticity creates a novel topological phase with
zero Chern number, in which a pair of topologically protected in-gap helical
edge states are found, protected by the 2D Zak phase associated fractional
polarization; this feature differs from the trivial phase with zero Chern
number and without edge states. Our findings provide insights into
symmetry-protected non-Hermitian topological insulators.

\section*{Appendix A: Energy band of the two-band equivalent Hamiltonian}

\label{Appendix A} 
\begin{figure}[t]
\includegraphics[ bb=0 0 520 525, width=8.7 cm, clip]{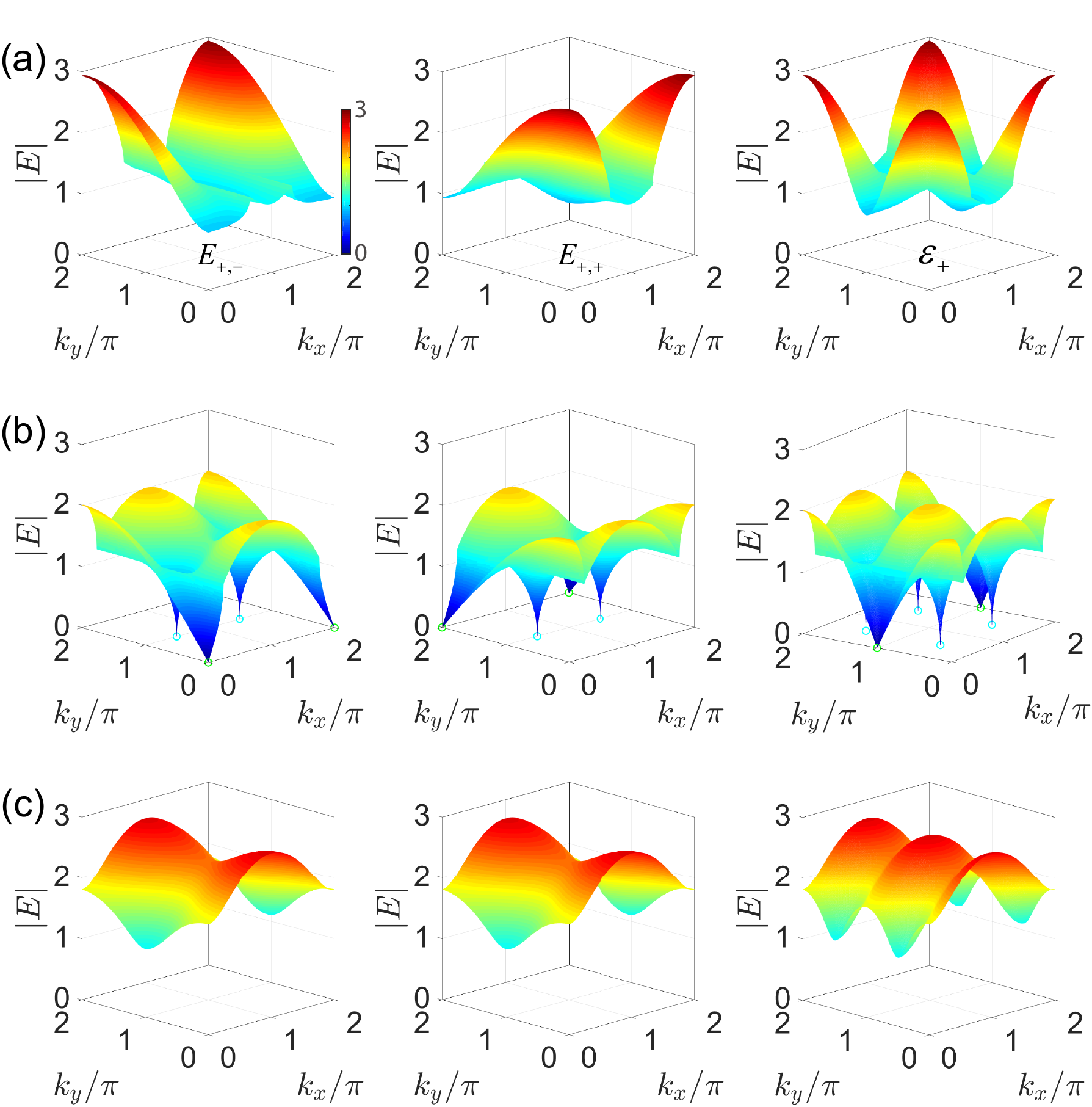}
\caption{Two upper bands $E_{+,\pm}$ of $\mathcal{H}(\mathbf{k}) $ are
depicted in the left and middle panels; the upper band $\protect\varepsilon_{+}$ of $h(\mathbf{k})$ is depicted in the right panel. The parameters are
identical to Figs.~\protect\ref{fig2}(a), (d), and (f). $m=t$; (a) $\protect\gamma=1/2$, (b) $\protect\gamma=\protect\sqrt{3}$, and (c) $\protect\gamma=5/2$. The EPs (DPs) are marked by the cyan (green) circles.}
\label{fighk}
\end{figure}

We show the energy band of the equivalent two-band Bloch
Hamiltonian $h(\mathbf{k})$ in Fig.~\ref{fighk} as the comparison with the
energy bands of the four-band Bloch Hamiltonian $\mathcal{H}(\mathbf{k})$
depicted in Fig.~\ref{fig2}. Three typical phases are presented. In Fig.~\ref%
{fighk}(a), we show the topologically nontrivial phase with $C=1$ to compare
with the energy bands depicted in Fig.~\ref{fig2}(a). In Fig.~\ref{fighk}%
(b), we show the gapless phase to compare with the energy bands depicted in
Fig.~\ref{fig2}(d). In Fig.~\ref{fighk}(c), we show the topologically
trivial phase with $C=0$ to compare with the energy bands depicted in Fig.~%
\ref{fig2}(f).

\section*{Appendix B: Robustness of in-gap edge state}

\label{Appendix B}

\begin{figure}[tbh]
\includegraphics[ bb=0 0 315 160, width=8.7 cm, clip]{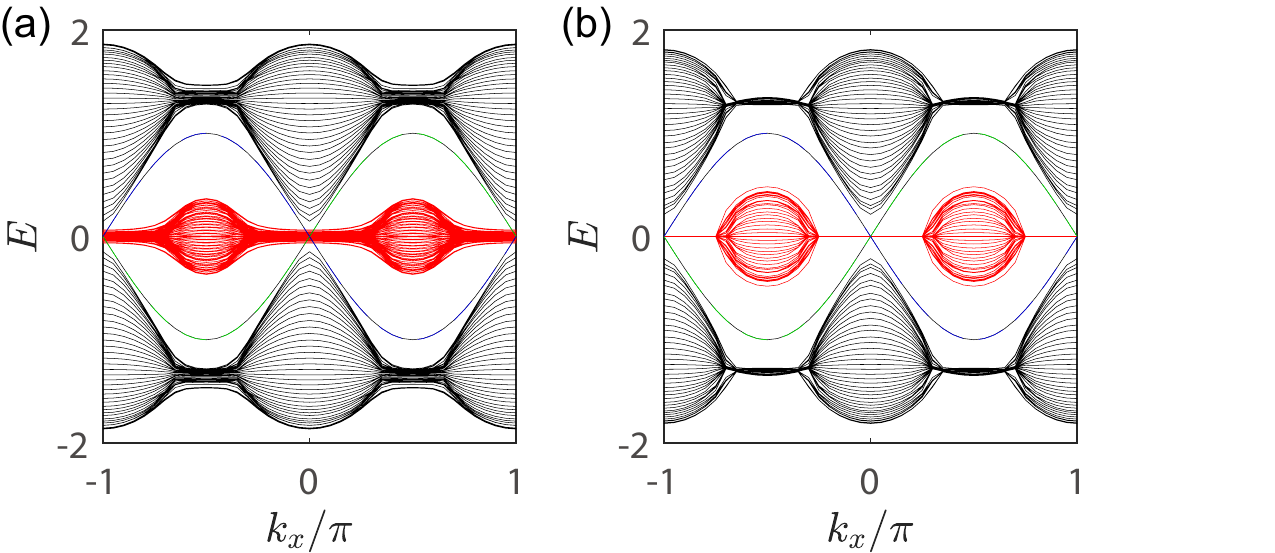}
\caption{Energy spectra under perturbation for the non-Hermitian Chern
insulator under OBC in the $y$ direction [Fig.~\protect\ref{fig5}(b)]. The
real (imaginary) part of $E$ is in black (red). The dashed green and blue
lines are the edge state energies $\pm t\sin k_{x} $ without perturbation.
The inversion symmetric perturbations are (a) $V_{a_{j}}=V_{d_{j}}=-V_{b_{j}}=-V_{c_{j}}=0.2R_j$ and (b) $V_{a_{j}}=V_{d_{j}}=-V_{b_{j}}=-V_{c_{j}}=0.2iR_j$, where $R_j$ is a random
real number within region $[0,1]$. Other system parameters are identical to
those in Fig.~\protect\ref{fig6}(e), $m=0$, $t=1$, and $\protect\gamma =1/2$. The number of unit cells of the 1D projection lattice is twenty.}
\label{figingap}
\end{figure}

We consider the 2D Chern insulator under OBC in the $y$ direction as
schematically illustrated in Fig.~\ref{fig5}(b). The in-gap edge states are
robust to the inversion symmetric perturbations. We demonstrate the
robustness to the inversion symmetric perturbations in the form of on-site
potentials (detunings), as well as gain and loss, respectively. The
inversion symmetry requires $V_{a_{j}}=V_{d_{j}}$ and $V_{b_{j}}=V_{c_{j}}$;
the pseudo-Hermiticity holds under $V_{a_{j}}+V_{b_{j}}=0$ and $%
V_{c_{j}}+V_{d_{j}}=0$. The energy spectra of the lattice under
perturbations are depicted in Fig.~\ref{figingap} for the comparison with
the energy spectrum in the upper panel of Fig.~\ref{fig6}(e). In Fig.~\ref%
{figingap}(a), the perturbations are the inversion symmetric on-site
potentials $V_{a_{j}}=V_{d_{j}}=-V_{b_{j}}=-V_{c_{j}}=0.2R_{j}$, where $%
R_{j} $ is a random real number within the region $\left[ 0,1\right] $ for
each unit cell $j$. In Fig.~\ref{figingap}(b), the perturbations are the
inversion symmetric gain and loss $%
V_{a_{j}}=V_{d_{j}}=-V_{b_{j}}=-V_{c_{j}}=0.2iR_{j}$. In both cases of Fig.~%
\ref{figingap}, the in-gap edge state energies are unchanged and remain in
the bandgap, being robust to the inversion symmetric perturbations.

\section{Acknowledgments}

This work was supported by National Natural Science Foundation of China
(Grants No.~11975128, No.~11605094 and No.~11874225), and the Fundamental
Research Funds for the Central Universities, Nankai University (Grants
No.~63191522 and No.~63191738).

\end{document}